# Microwave vortex beam lasing via photonic time crystals

Lei Huang[1†], Weixuan Zhang[1†*], Deyuan Zou[1], Jiacheng Bao[2], Fengxiao Di[1], Haoyu Qin[1], Long Qian[1], Houjun Sun[2*], Xiangdong Zhang[1*]

[1]*Key Laboratory of advanced optoelectronic quantum architecture and measurements of Ministry of Education, Beijing Key Laboratory of Nanophotonics & Ultrafine Optoelectronic Systems, School of Physics, Beijing Institute of Technology, 100081 Beijing, China.*
[2]*Beijing Key Laboratory of Millimeter wave and Terahertz Techniques, School of Information and Electronics, Beijing Institute of Technology, Beijing 100081, China*
[†]*These authors contributed equally to this work.*
*Corresponding author. E-mail: zhangxd@bit.edu.cn; sunhoujun@bit.edu.cn; zhangwx@bit.edu.cn

**Microwave lasing carrying orbital angular momentum (OAM) holds significant potential for advanced applications in fields such as high-capacity communications, precision sensing, and radar imaging. However, conventional approaches to masers fail to produce emission with embedded OAM. The recent emergence of photonic time crystals (PTCs)—artificially structured media with periodically varying electromagnetic properties in time—offers a paradigm shift toward resonance-free lasing without the need for gain media. Yet, pioneering PTC designs have been based on three-dimensional bulk structures, which lack a surface-emitting configuration, and do not possess the capability to modulate OAM, thus hindering the realization of surface-emitted PTC masing that carries OAM. Here, we report the first experimental demonstration of non-resonant, gain medium-free, and surface-emitted microwave vortex beam lasing OAM using ring-shaped PTCs. By developing a multiplier-driven time-varying metamaterial that achieves over 100% equivalent permittivity modulation depth, we establish momentum bandgaps ($k$-gaps) with sufficient bandwidth to overcome intrinsic losses and enable self-sustained coherent microwave amplification. Furthermore, space-time modulation induces non-reciprocity between clockwise and counterclockwise $k$-gap modes within the circularly symmetric PTC structure, facilitating the selective generation of microwave lasing carrying OAM—a capability beyond the reach of conventional maser technologies. Our work bridges PTC physics with coherent OAM-carrying microwave emission, establishing a transformative platform for next-generation wireless communications, advanced sensing systems, and OAM-based technologies.**

Time-varying media have recently emerged as a transformative paradigm across wave physics, spanning quantum mechanics, photonics, acoustics, and thermal science[1–8]. Unlike static systems, these dynamically modulated materials exhibit extraordinary phenomena and enable functionalities that are fundamentally unattainable in conventional platforms. Among them, photonic time crystals (PTCs)[9–15]—characterized by periodic temporal modulation of electromagnetic parameters such as permittivity—have garnered significant interest. Their unique interplay of temporal reflections and refractions generates anomalous dispersion bands featuring momentum bandgaps ($k$-gaps), enabling exotic phenomena like momentum-selective dynamical gain[15], topological phase transitions in $k$-gaps[16–18], free-electron radiation amplification[19], and others[20–30]. Critically, PTCs theoretically support cavity-free lasing via mirrorless and non-resonant amplification[15]—a revolutionary departure from cavity-dependent lasers that could transcend traditional gain-medium limitations. However, translating this principle into practice faces formidable challenges, particularly at optical frequencies where prohibitive modulation speeds and amplitudes are required.

Meanwhile, masers[31]—the microwave precursors to lasers[32-35]—persist in specialized applications such as deep-space communications and quantum technologies. Their broader impact remains constrained by inherent limitations: conventional designs rely on high-Q resonant cavities and restrictive gain-media energy structures, leading to narrow frequency tunability and suboptimal energy efficiency. Operational requirements for low temperatures, high-vacuum environments, or intense magnetic fields[36-42] further hinder practical deployment. Moreover, traditional masing schemes have so far been confined to the emission of unstructured fields and lack advanced functionalities—such as the generation of microwave lasing carrying orbital angular momentum (OAM). This raises a pivotal question: Can PTC physics bypass these constraints to enable room-temperature, surface-emitted and gain-medium-free maser-like emission with OAM? In fact, realizing this PTC-driven microwave lasing also faces great difficulties. A pioneering work experimentally reveals the non-Hermitian band structure of photonic Floquet media by implementing a time-varying capacitance system using varactor diodes inside a closed metallic waveguide[28]. In this system, noise-initiated microwave oscillations originating from k-gaps in PTCs have been explicitly observed. However, this work is based on closed structures, which inherently prevent out-of-plane emission and the generation of structured masing fields. In contrast, an open-structure PTC system using varactor diodes has demonstrated microwave amplification[30]. Nevertheless, due to the limited modulation strength of varactor diodes, the surface-emitted microwave lasing has not yet been achieved. This raises an important question: can one design an open PTC system that exploits $k$-gaps to achieve surface-emitted microwave lasing, while also enabling structured-light emission carrying OAM?

In this work, we overcome these challenges through a new type of spatiotemporal metamaterial, achieving the first experimental realization of surface-emitted microwave vortex beam lasing in PTCs. We develop a multiplier-driven time-varying capacitance architecture that delivers an unprecedented 100% equivalent permittivity modulation depth, activating robust momentum bandgap amplification. Full-wave simulations and microwave experiments demonstrate threshold-governed maser-like emission from PTCs in both near-field and far-field regions. Notably, space-time modulation in our system induces symmetry breaking that generates vortex-beam masing carrying OAM—a capability unattainable in conventional masers. Our work establishes a unified framework integrating PTC physics with microwave engineering, creating a new paradigm for active metamaterials applicable to deep space communication, hyperspectral sensing, and OAM-enabled signal processing.

**Theory of surface-emitted microwave lasing carrying OAM via spatiotemporally engineered photonic time crystals.** Pioneering works on PTC-based lasing have primarily investigated spatially infinite structures[15], where the permittivity oscillates periodically in time while maintaining spatial uniformity in all three dimensions. However, implementing such spatially extended PTCs is experimentally challenging. Moreover, the lack of angular spatial modulation prevents these systems from generating radiation with OAM. Here, we design a ring-shaped PTC confined in a two-dimensional surface to overcome these limitations. Figure 1a illustrates the conceptual design: a ring-shaped spatiotemporally modulated surface capacitance with radius $R$, which enables surface-emitting microwave lasing with OAM in the vertical direction. Specifically, the spatiotemporal surface capacitance $C(\phi, t) = C_0[1 + \delta cos(q\phi - \Omega t)]$ (illustrated in the bottom chart of Fig. 1a) exhibits the rotational space-time symmetry $C(\phi, t) = C\left(\phi + \theta, t + \frac{q\theta}{\Omega}\right)$, the time periodicity $C(\phi, t) = C\left(\phi, t + \frac{2\pi}{\Omega}\right)$ and the angular periodicity $C(\phi, t) = C\left(\phi + \frac{2\pi}{q}, t\right)$, where $(\theta, \frac{q\theta}{\Omega})$ is the space-time rotational vector, $\phi \in [0, 2\pi)$ is the angular coordinate, $\Omega$ is the modulation frequency and integer $q$ governs the angular periodicity. Under these symmetries, we can expand $C(\phi, t)$ as a Fourier series $C(\phi, t) = \sum_m c_m e^{im(q\phi - \Omega t)}$, where $c_m$ are the Fourier coefficients ($m \in Z$). Here, we assume that transverse electric eigen-waves propagate along the $\phi$ direction in this configuration. Thus, the electric field ($E_\rho$) is radially polarized, and the magnetic field comprises azimuthal ($H_\phi$) and vertical ($H_z$) components. Based on the generalized Floquet-Bloch theory, the eigen-waves in cylindrical coordinates take the form:

$$\begin{cases} E_\rho = \sum_n E_{\rho,n} e^{-\gamma_n z} e^{i[(p+nq)\phi-(\omega+n\Omega)t]}, \\ H_\phi = \sum_n H_{\phi,n} e^{-\gamma_n z} e^{i[(p+nq)\phi-(\omega+n\Omega)t]}, \\ H_z = \sum_n H_{z,n} e^{-\gamma_n z} e^{i[(p+nq)\phi-(\omega+n\Omega)t]}, \end{cases} \quad (1)$$

where $\gamma_n$ represents the radiation rate of eigen-waves toward background. The $p$ is related to the wavevector $k$ by $p = kR$, and $\omega$ defines the eigen-frequency. Substituting Eq. (1) into free-space Maxwell's equations and surface-capacitance boundary conditions, we obtain two coupled relations (see S1 of the Supplementary Materials for the detailed derivation):

$$\begin{cases} H_{\phi,n} = \sum_m -i(\omega+n\Omega) E_{\rho,m} c_{n-m}, \\ H_{\phi,n} = \dfrac{\sqrt{\dfrac{(p+nq)^2}{R^2} - \dfrac{(\omega+n\Omega)^2}{c^2}}}{i\mu_0(\omega+n\Omega)} E_{\rho,n}. \end{cases} \quad (2)$$

These equations are reformulated into a matrix eigenvalue problem, $\mathbf{YE} = \mathbf{H}$ and $\mathbf{ME} = \mathbf{H}$, where $\mathbf{Y}$ and $\mathbf{M}$ encode the surface capacitance modulation and free-space interactions between $\mathbf{E}$ and $\mathbf{H}$, respectively. Solving $\det(\mathbf{Y} - \mathbf{M}) = \mathbf{0}$ reveals the eigen-spectrum of the ring-shaped surface-capacitance PTC model.

Figures 1b-c present calculated eigen-spectra for temporal ($q$=0) and spatiotemporal ($q$=1) modulations, respectively. Here, the system parameters are set as $R$=0.4 m, $\frac{\Omega}{2\pi}$=634 MHz, $C_0$=2 pF and $\delta$=0.5. We find that the ring-shaped PTC possesses equally-spaced discrete eigenmodes due to the angular periodicity (pink and green dots correspond to real and imaginary eigen-frequencies), where the system with a larger circumference possesses more discrete eigenmodes. It is shown that both temporal and spatiotemporal modulations can introduce $k$-gaps (marked by shaded areas), which sustain gain modes enabling coherent wave amplification. For comparison, we also plot the band dispersion of a spatial infinite PTC characterized by $C(x,t) = C_0[1 + \delta cos(Kx - \Omega t)]$ with $K = q/R$, as shown by solid lines in Figs. 1b-c. It is seen that discrete eigenmodes of our ring-shaped PTC are perfectly embedded within the band dispersion of the spatial infinite PTC, showing the effectiveness for implementing PTC using ring-shaped geometry. In addition, it is worth noting that, under temporal modulation ($q$=0), the system preserves continuous rotational symmetry, resulting in identical dispersion bands for clockwise (CW, $kR$>0) and counterclockwise (CCW, $kR$<0) propagating eigen-waves. In contrast, spatiotemporal modulation ($q$=1) breaks the reciprocity, yielding asymmetric dispersion curves for opposite wavevectors. This asymmetry confirms the nonreciprocal nature of the space-time modulated system, where CW and CCW $k$-gap eigen-waves exhibit distinct frequencies ($\omega_1 \neq \omega_2$ and $\omega_3 \neq \omega_4$).

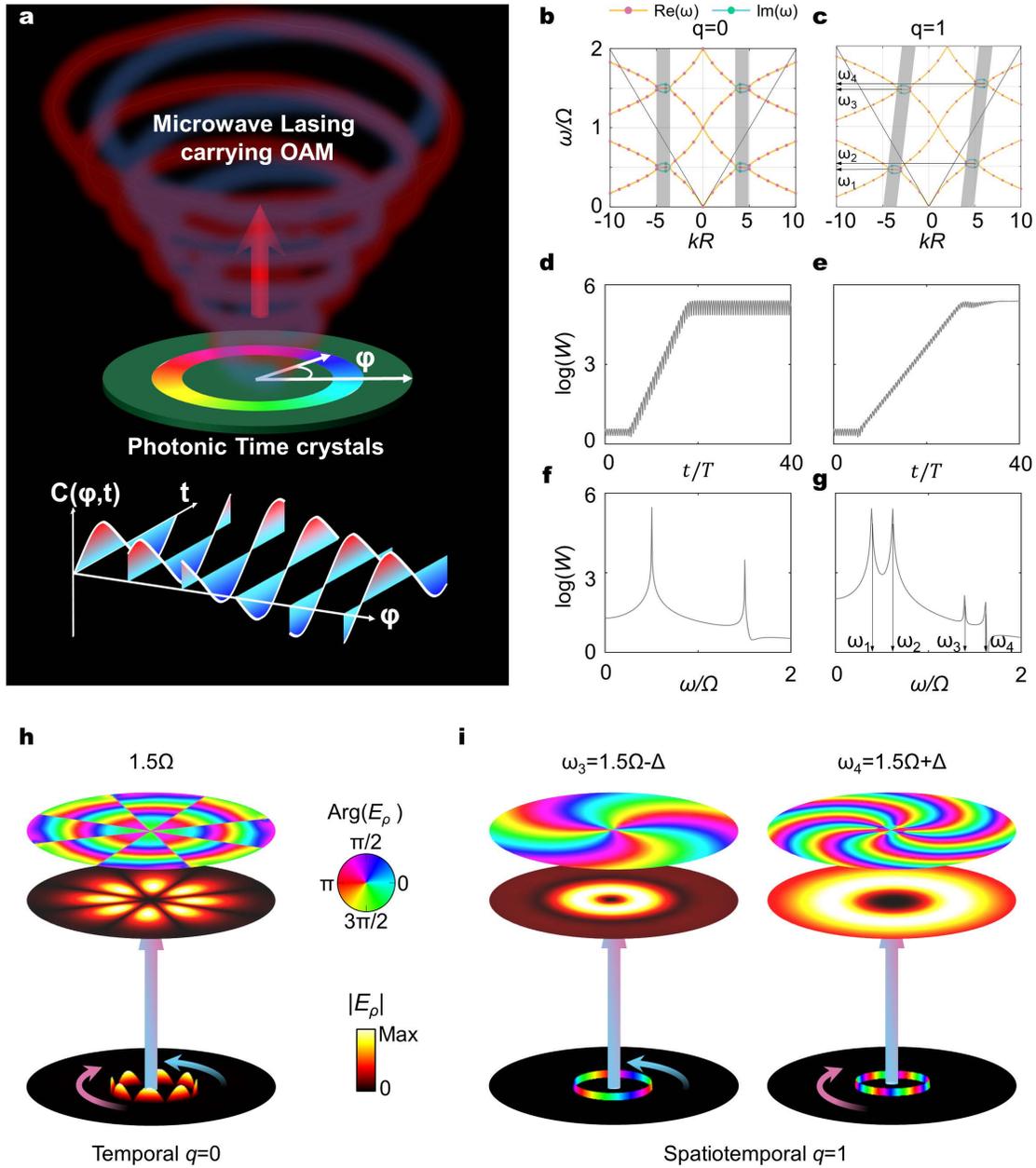

**Fig. 1 | Theoretical results of surface-emitted microwave lasing carry OAM via spatiotemporally engineered photonic time crystals. a,** Schematic of the surface-emitted masing with OAM from ring-shaped spatiotemporally modulated PTC. Colored gap: space-time modulated surface capacitance $C(\phi, t)$. **b-c,** Eigenenergy spectra for temporal ($q=0$, (b)) and spatiotemporal ($q=1$, (c)) modulation. Dots: discrete eigenmodes of the ring-shaped PTC. Yellow curves correspond to the spatially infinite PTC $C(x,t)$. Asymmetric $k$-gap bands ($q=1$) confirm nonreciprocity. **d-g,** Time-domain evolution (d and e) and Fourier spectra (f and g) of near-field responses under $q=0$ (d and g) and $q=1$ (e and g). Saturation effects and absorption losses are included. Spectral splitting ($q=1$) corresponds to nondegenerate forward/backward modes. **h-i,** Far-field amplitude and phase distributions of dominant frequency components. The uniform intensity and zero OAM for $q=0$ in (h). The spiral phase fronts (-3 and +6 topological charges for two frequencies) confirm OAM generation for $q=1$ in (i). Other parameters are set as: ring radius $R=0.4$

m, $\Omega/2\pi = 634$ MHz, and capacitance profile $C(\phi, t) = C_0[1 + \delta cos(q\phi - \Omega t)]$, $C(x, t) = C_0[1 + \delta cos(Kx - \Omega t)]$ with $K = q/R$, $C_0 = 2$ pF and modulation depth $\delta = 0.5$.

To further illustrate the microwave lasing in ring-shaped PTCs, Figs. 1d and g depict time-domain evolutions and Fourier spectra of a fixed near-field point under temporal and spatiotemporal modulations, respectively. Here, the time-domain simulation is divided into two segments: from $t=0$ to $5T$ ($T = \frac{2\pi}{\Omega}$), a single-frequency source $I(t)=\sin(\Omega t)$ is used to excite the system under time-invariant surface capacitance ($\delta=0$), and in the range of $t>5T$, the source is removed and time-varying capacitance activates to observe the field evolution under $k$-gap. In addition, to approximate realistic conditions, we incorporate the saturation effect of time-varying capacitance in our simulations (the modulation depth $\delta$ decreases with increasing field intensity, see details in Methods I). For the temporal modulation ($q=0$), the initial exponential growth of the electric field—driven by $k$-gap gain exceeding absorption losses—appears and gradually saturates into a steady state. The Fourier spectrum of the stabilized field reveals eigenmodes at odd multiples of $0.5\Omega$, precisely aligned with the $k$-gap modes in Fig. 1b. Under spatiotemporal modulation ($q=1$), the field similarly exhibits $k$-gap enabled exponential growth followed by saturation. While, different from the case with $q=0$, the Fourier spectrum splits into nondegenerate peaks ($0.5n\Omega \pm \Delta$, $n$ is odd number) in each Floquet sector, corresponding to CW and CCW $k$-gap eigenmodes. This is consistent with the asymmetric band structure with respect to $k$-vector in Fig. 1c.

Then, we investigate the far-field radiation characteristics of surface-emitted masing in both temporal and spatiotemporal PTCs. The far-field radiation is calculated from the near-field distribution using angular spectrum theory (details provided in Method II). Figure 1h depicts the near-field intensity distribution (lower panel) and the far-field phase and amplitude distributions (two upper panels) of the $E_\rho$-component at $1.5\Omega$ under temporal modulation. The near-field intensity distribution exhibits a standing-wave pattern due to the simultaneous excitation of degenerate CW and CCW modes. Consequently, as this pair of degenerate modes generates OAM beams with opposite topological charges in the far field, the total far-field radiation carries net zero OAM. In contrast to the purely temporal case, spatiotemporal modulation can break reciprocity of CW and CCW modes, inducing nondegenerate eigen-frequencies with opposite propagation directions in each Floquet sector. Two charts in Fig. 1i present the calculated near- and far-field distributions for the spatiotemporally modulated case at $1.5\Omega - \Delta$ (CCW) and $1.5\Omega + \Delta$ (CW), respectively. It is shown that far-field phase profiles display spiral structures with +6 and -3 topological charges related to CW and CCW modes at $1.5\Omega + \Delta$ and $1.5\Omega - \Delta$, unambiguously confirming the generation of coherent microwave emission carrying OAMs.

In S2 of the Supplementary Materials, we further explore the impact of ring circumference on the divergence angle of PTC-based microwave lasing, demonstrating size-enhanced directivity due to reduced diffraction effects. It is crucial to emphasize that, in contrast to conventional approaches relying on high-Q whispering-gallery-mode resonances to achieve lasing through strong light-gain medium interactions, our masing mechanism eliminates the need for a resonant cavity and gain medium. Instead, the self-sustained amplification arises entirely from the *k*-gap modes of spatiotemporally PTC and the ring-shape geometry is used to enable vertical radiation and CW/CCW eigenmodes with OAMs.

Furthermore, it is worth noting that the although mode-locked laser system also contains a ring configuration and time modulation, there are fundamental differences between it and our system. In mode-locked lasers, a gain medium is present, and the modulation signal is mainly used to lock the phases of different longitudinal modes. Importantly, the modulation frequency is much lower than the lasing frequency, which allows the system to satisfy the slow-envelope approximation and to be described by a Schrödinger-type equation along the propagation direction[44-48]. In contrast, our system does not include any gain medium, and the modulation frequency exceeds the fundamental masing frequency. Therefore, it must be described directly by Maxwell's equations. The detailed comparison is provided in S3 of the Supplementary Materials.

**Practical implementation of surface-emitted microwave lasing carrying OAM by spatiotemporal metamaterials.** To experimentally realize surface-emitted microwave lasing with OAMs in ring-shaped PTCs, in this part, we design a microwave time-varying metamaterial that effectively emulates the spatiotemporally modulated surface-capacitance model described above. Conventional approaches for microwave time-varying metamaterials rely on the capacitance-tuning effect of varactor diodes[12,28–30], where equivalent capacitance varies with applied modulation voltage. However, the achievable capacitance modulation depth is severely restricted by the physical properties of varactor diodes. Specifically, under a reverse direct voltage bias of 3.5 V, applying a 1V alternating voltage modulation yields a capacitance variation of only 10% (see S4 of the Supplementary Materials for comparative between varactor diodes and multiplier-driven capacitors). Furthermore, higher modulation voltages drive diodes into the forward-bias region, causing failure. These factors critically constrain the temporal modulation capability—manifested as insufficient momentum bandgap width and limited microwave gain—rendering the system hard to overcome losses and generate self-sustained coherent microwave emission.

To overcome these limitations of varactor-based designs, we develop a new multiplier-driven time-varying capacitance metamaterial, as depicted in Fig. 2a. A ring-shaped gap (the green region with radius $R$ and width $g$) is etched into the top metal plane (orange regions) to form a planar waveguide, which is coupled with a few numbers of multiplier-driven time-varying capacitor (as highlighted by black labels). Figure 2b shows the cross section of the structure, illustrating the integration between the planar waveguide and the time-varying capacitive circuit. Specifically, in the multiplier-driven time-varying capacitor, the microwave multiplier ADL5391 receives an external modulation signal $V_n(t)$ ($n = [1,6]$) at one input port and the other couples to the planar waveguide through the right via. The output port of multiplier connects in series with a fixed capacitor $C_0$ and feeds back to the planar waveguide through the left via. Leveraging the multiplier function W =$\alpha$XY (W is the output. X and Y are inputs. $\alpha$ is the multiplier factor with $\alpha$=1 used), we achieve an effective time-varying capacitance $C = C_0[1 - V(t)]$ (see derivations in Methods III). Compared to varactor-based designs, this multiplier-driven approach offers three key advantages: 100% modulation depth at $|V(t)|$ =1 V (enabling larger $k$-gaps), the linear modulation under high voltages (free from diode nonlinearities), and the signal isolation between modulation and propagation paths (eliminating filter requirements). To validate the effectiveness of our design, we perform full-wave electromagnetic simulations on a small-scale planar waveguide incorporating a single multiplier-driven capacitance unit (see S5 of the Supplementary Materials for numerical results). In these simulations, the multiplier and $C_0$ are modeled as lumped circuit elements, while the vias and feedlines are simulated with their actual geometries to accurately account for structural losses and dispersion effects (see Methods I for simulation details). Crucially, unlike the idealized surface-capacitance model, the implemented time-varying ring-shaped planar waveguide exhibits significant frequency dispersion, which constrains the operational bandwidth of PTCs with a high transmittivity (see S5 of the Supplementary Materials for the transmission spectra of static PTC metamaterials). Therefore, we must design structural parameters to ensure that the fundamental $k$-gap eigenmode at 0.5Ω resides within the static planar waveguide's bandpass region. Under this condition, our designed multiplier-driven capacitance metamaterial can generate a large $k$-gap, leading to coherent microwave amplification.

To implement the spatiotemporal PTCs, we integrate six multiplier-driven time-varying capacitance modules around the ring-shaped planar waveguide (Fig. 2a), which is equivalent to the discretized surface-capacitance model shown in Fig. 2c. Here, each distinctively colored segment with $\phi \in [\phi_s + n\pi/3, \phi_f + n\pi/3]$ ($n \in [1, 6]$) corresponds to a uniform time-varying capacitive region associated with one multiplier module. The different colors represent $C_n(t) = C_p + C_{mod}\sin(2\omega t + \varphi_n)$ with distinct phases $\varphi_n$, designed to match the phase $q\phi$ of the continuous spatiotemporal modulation model within each discrete angular sector. Between each pair of time-varying capacitor modules, the static surface capacitance $C_s$ is applied, representing the inherent static capacitance of the planar waveguide segments connecting the modulated modules. Using the finite element simulation, we determine effective values of $C_p$, $C_{mod}$, $C_s$, $\phi_s$ and $\phi_f$ for the time-varying and static effective capacitances of the PTC structure (see S6 of the Supplementary Materials for simulation results and parameter values). These values are then used to define the discretized spatiotemporal surface capacitance models for full-wave simulations.

The simulated wave evolutions and frequency spectra under temporal modulation ($\varphi_n = 0$) at near-field and far-field points are shown in Figs. 2d-e and Figs. 2f-g, respectively. Other parameters are presented in the figure caption. It is shown that both near-field and far-field waves exhibit exponential growth (highlighted by the blue regions) under time-varying capacitance modulation, saturating to a steady state due to the multiplier output limit. Spectral peaks appear at ω, 3ω, 5ω, which match to *k*-gap eigen-frequencies in the continuous counterpart. Notably, while the ideal surface-capacitance model (Fig. 1a) places the 3ω-component eigenmode below the light line (non-radiative), the discretized capacitance distribution transforms the continuously rotational symmetry to discrete rotational symmetry, which can couple the 3ω-mode's wavevector above the light line[49], as evident by the larger far-field radiation intensity at 3ω than that at ω (see S7 of the Supplementary Materials for details on how spatial discretization couples modes above the light cone).

Finally, we investigate the masing behavior in the spatiotemporal PTC metamaterial by configuring the modulation phase as $\varphi_n = n\pi/3$ ($n \in [1,6]$). The calculated near-field (Fig. 2h) and far-field (Fig. 2j) temporal evolutions again show pronounced exponential growth (highlighted by blue regions) followed by saturation. The corresponding frequency spectra (Figs. 2i and k) reveal that the spectral splitting occurs in each Floquet sector. The oscillation feature observed in Fig. 2i and k originates from the discretized spatiotemporal modulation used in the simulation. This splitting behavior is consistent with the predictions of the continuous surface-capacitance model (Fig. 1g), although its magnitude is reduced compared to the continuous case. It is important to emphasize a key distinction from the continuous case: in the continuous limit, CW and CCW *k*-gap modes are fully decoupled. However, the spatial discretization in the practical design introduces the coupling between these two eigen-modes. Consequently, the two split frequency components no longer correspond purely to ideal CW and CCW eigenmodes in the continuum limit, but rather represent superposition of both modes. The coupling coefficients governing this superposition depend on factors including the density of the time-varying capacitor modules, the circumference of the ring PTC, and the impedance of PTCs (see S8 of the Supplementary Materials for the parameter dependence of the effective couplings).

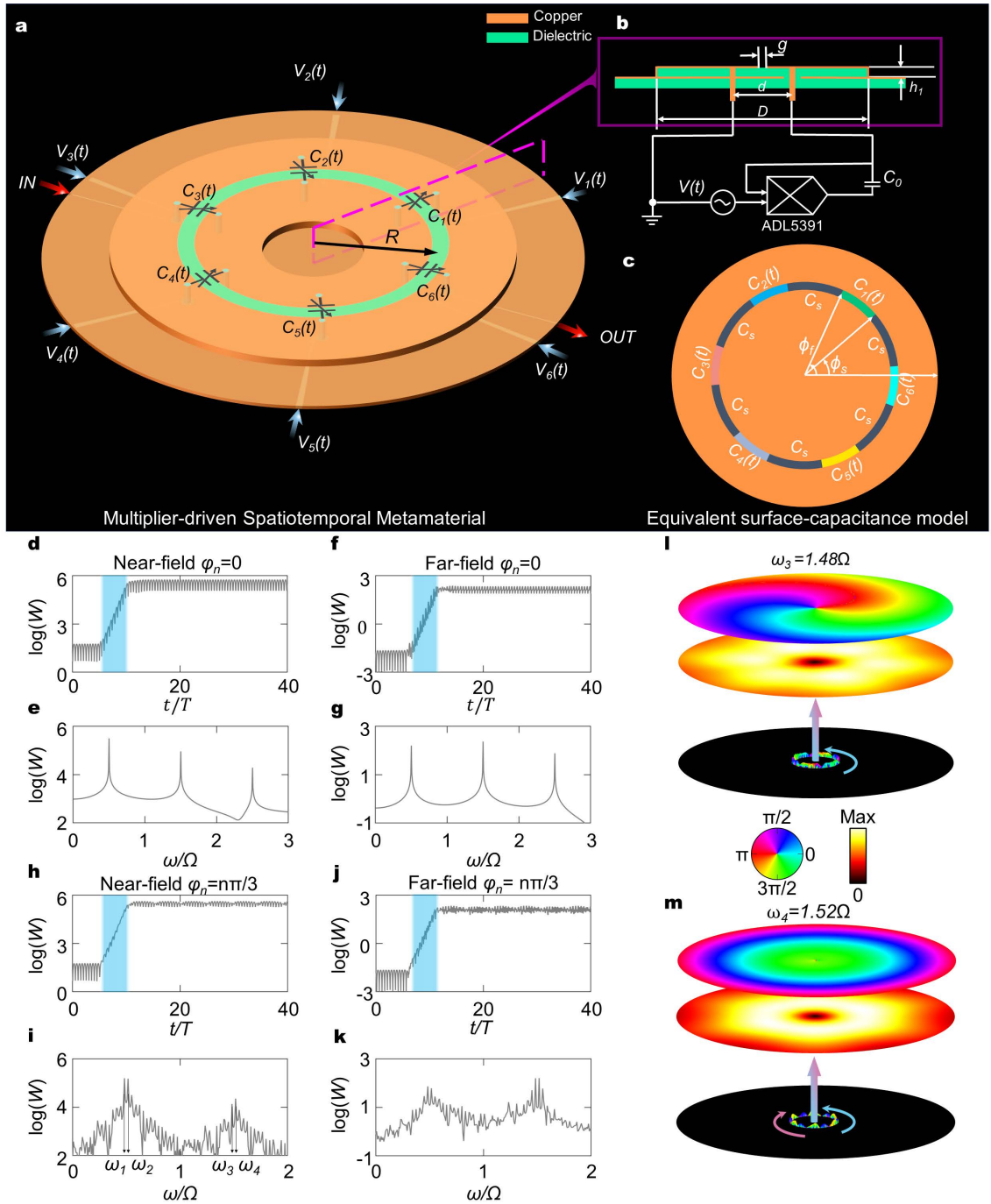

**Fig. 2 | Multiplier-driven spatiotemporal metamaterial and surface-emitted microwave lasing with OAMs.** PTC metamaterial design: **a**, Top-view schematic of ring-shaped planar waveguide structure with six multiplier-driven time-varying capacitance units. **b**, Cross-section showing dielectric substrate (thickness $h_1$), ground plane, and control circuit (inset: multiplier ADL5391 with fixed capacitor $C_0$=10 pF). Key parameters: ring radius $R$=0.1 m, gap width $g$=1 mm, via spacing $d$=9 mm. **c**, The schematic diagram of the discretized surface-capacitance model for the multiplier-driven spatiotemporal metamaterial. **d-k**, Temporal modulation results ($\varphi_n$=0): (d and e) Near-field intensity evolution (d) and spectrum (e). (f and j). Far-field counterparts showing

fundamental radiation at ω. (h-k). Spatiotemporal modulation ($\varphi_n=n\pi/3$): (h-i), Near-field and (j-k), far-fiel. Spectral splitting (22.8 MHz) evident in insets. **l-m**, Reconstructed far-field phase profiles based on angular spectrum theory at split frequencies confirming 1 OAM states. Scale bars: phase from 0 to 2π. Simulation parameters: $\omega/2\pi$=304 MHz, $\Omega/2\pi$=608 MHz, $R$= 0.1 m, and $\alpha$=0.7.

To further characterize the OAM properties of two split eigenmodes in spatiotemporal PTCs, we compute far-field radiation patterns using angular spectrum theory based on the simulated near-field distributions at $\omega_1 = 1.48\Omega$ and $\omega_2 = 1.52\Omega$, as shown in Figs. 2l-m. It is shown that, due to the strong near-field coupling between CW and CCW modes at $\omega_2 = 1.52\Omega$, the non-chiral wave profile appears, preventing its OAM properties in the far field. In contrast, the CCW-dominated near-filed profile still exists at $\omega_1 = 1.48\Omega$, resulting in a helical phase with topological charge being -1 in the far-field region. It is worth noting that by suitably engineering the structural parameters, we can make two split eigenmodes in a fixed Floquet sector become dominated either by CW or CCW (see S8 of the Supplementary Materials for detailed simulation results), enabling a coherent microwave emission carrying OAMs. These numerical results unambiguously demonstrate the generation of surface-emitted coherent microwave amplification carrying OAMs in our PTC metamaterials—a capability not yet demonstrated in conventional maser systems relying on static cavities and gain media.

**Experimental observation of surface-emitted microwave lasing carrying OAM by spatiotemporal metamaterials.** Building on the theoretical framework, we fabricate the designed spatiotemporal metamaterials, featuring a ring-shaped planar waveguide with radius $R$=0.1 m, gap width $g$=1 mm and six multiplier-driven time-varying capacitance modules, as shown in Figs. 3a-b for the front and back views. The enlarged inset in Fig. 3a details a time-varying capacitance module with the ADL5391 multiplier is positioned centrally. It's first differential input port ($P_1$) connects to an SMA port via a 50 Ω feedline, with transformer (TC1-1-13M) in series to achieve impedance matching and single-ended-to-differential signal conversion. The second input port ($P_2$) connects to the planar waveguide. The output port of multiplier ($O$) connects to the planar waveguide in series with a 10 pF capacitor $C_0$. To suppress low-frequency crosstalk, all input/output and power interfaces are configured with filter capacitors (see Supplementary Table S2 for these capacitance values). Simultaneously, the multiplier's terminal is controlled by a switch (SW) to compare ON/OFF modulation conditions (see Methods IV for details on the sample fabrication).

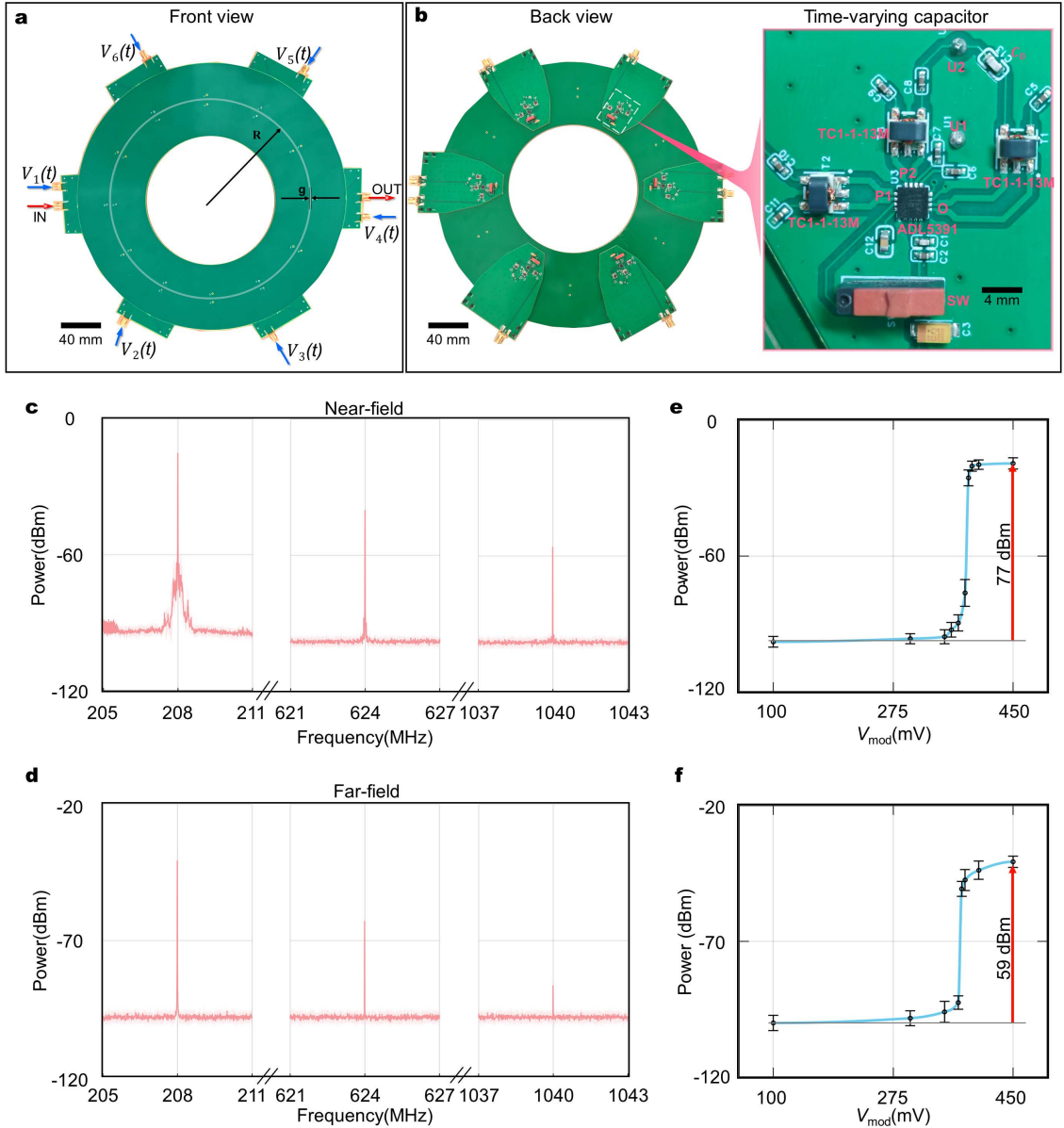

**Fig. 3 | Experimental results of temporal PTC-enabled microwave lasing. a-b**, Fabricated spatiotemporal metamaterial, with front(a) and back(b) views. Ring-shaped planar waveguide transmission line ($R$=0.1 m, $g$=1 mm) with six integrated multiplier-driven time-varying capacitance modules. Inset: Single module detail showing ADL5391 multiplier with: $P1$ – differential input port connected to SMA via 50Ω feedline and TC1-1-13M transformer (single-ended-to-differential conversion); $P2$ – input linked to planar waveguide through $U_1$ via; $O$ – output connected to $U_2$ via through 10 pF capacitor $C_0$. Filter capacitors suppress low-frequency crosstalk (values in Supplementary Table 1). **c**, The measured near-field spectrum at the output port. Dominant peak at 208 MHz (half the 416 MHz driving frequency) with other odd harmonics (624 MHz, 1040 GHz). The red curve shows the mean intensity and shaded regions manifest the fluctuation. **d**, Measured far-field spectrum at 1 m distance from the sample. Fundamental radiation peak at 208 MHz with 31 dB attenuation relative to near-field. **e-f**, Threshold-governed transitions of near-field and far-field steady-state intensities versus the modulation strength $V_{mod}$. Dashed line

marks lasing threshold at 0.37 V. Sub-threshold ($V_{mod}$ < 0.37 V): signals at noise floor. Super-threshold: surge to saturation.

At first, we apply the homogenous modulation signal, $V_n(t) = V_{mod}sin(2\omega t)$ with $\omega/2\pi$=208 MHz and $V_{mod}$=0.45 V, to all modulation ports (marked by blue arrows in Fig. 3a) to observe temporal PTC-enabled coherent microwave emission. We measure the wave evolution at the output port (the red arrow, Fig. 3a) and plot the corresponding frequency spectrum in Fig. 3c. It is shown that the frequency spectrum exhibits dominant peaks at $\omega$, $3\omega$ and $5\omega$, confirming that *k*-gap modes govern the wave evolution in the temporal PTC. Furthermore, we measure the frequency spectrum at a far-field point located at *z*=1 m from the PTC structure, as shown in Fig. 3d. It is shown that the ω-mode remains dominant in the far-field radiation, being different from the 3ω-mode dominated far-field emission shown in Fig. 2g. This is attributed to the breaking of $C_6$ symmetry in the fabricated sample (see S9 of the Supplementary Materials for details). Moreover, it is worth noting that the modes at 3ω and 5ω also exhibit surface-emitted masing effects. This indicates that high-frequency masing can be generated using a low-frequency modulation signal in our PTC metamaterials. Crucially, this high-frequency signal generation originates fundamentally from the temporal modulation effect, which is distinctly different from conventional material-based nonlinearities[50].

To further investigate the threshold behavior of microwave lasing in PTCs, we characterize the steady-state electric field intensities versus the modulation voltage $V_{mod}$ for both near-field (Fig. 3e) and far-field (*z*=1 m, Fig. 3f) regions. When $V_{mod}$<0.37 V, both intensities remain below the detection limit. This indicates a loss-dominant regime after removing the excitation signal, where the PTC *k*-gap is insufficient to overcome intrinsic losses (e.g. the absorption and radiation losses) in the system and cannot initiate microwave lasing. For $V_{mod}$>0.37 V, both near-field and far-field intensities surge to saturation levels, demonstrating the self-sustained microwave lasing in PTCs. This transition confirms the *k*-gap amplification surpassing intrinsic losses. Our experimental results establish that the multiplier-driven time-varying metamaterial achieves sufficient equivalent permittivity modulation to open large *k*-gaps to overcome intrinsic losses, enabling the self-sustained coherent microwave amplification.

Lastly, we implement the spatiotemporally modulated PTCs by configuring the phases of six modulation signals as $\varphi_n = n\pi/3$ ($n \in [1,6]$). All other parameters remained identical to the temporal PTC case. Figures 4a-b display the measured frequency spectra at near-field (the output port) and far-field (*z*=1 m) locations with $V_{mod}$=0.5 V, respectively. It is found that both of two frequency spectra exhibit significant frequency-splitting behavior: the single peak observed in temporal PTC ($\varphi_n = 0$) splits into dual-frequency curves. This phenomenon aligns with the full-wave simulation of the discretized surface-capacitance model. Moreover, it is worth noting that the measured static transmission spectrum of our sample shows a distinct peak in a narrow frequency range (see Fig. S4). Consequently, the frequency components in the low-transmission region are accompanied by strong propagation loss and are buried in the background noise, rendering the oscillatory feature (shown in Fig. 2I) unobservable in the experimental measurements. Figures 4c-d present the steady-state intensities at near-field and far-field points (consistent to Figs. 4a-b) versus the external modulation strength $V_{mod}$. We find that, below the microwave lasing threshold, field intensities remain minimal below the detection limit, confirming loss-dominated behavior in the system. While, when $V_{mod}$ exceeds the threshold of 0.44 V, the steady-state intensities exhibit step-like jumps. These results demonstrate that the $k$-gap amplification can overcome the system losses and induce self-sustained coherent masing in spatiotemporal PTCs.

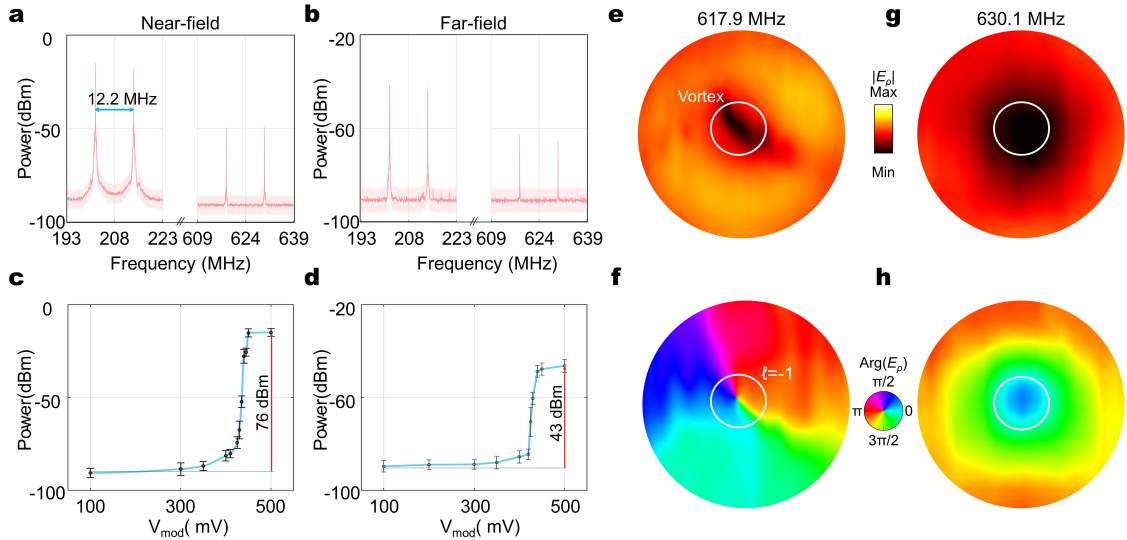

**Fig. 4 | Observation of microwave lasing with OAMs via spatiotemporal PTCs. a**, Near-field spectrum of spatiotemporal PTC with non-reciprocal propagation effects. Frequency splitting from 208 MHz (temporal PTC) to dual peaks at $f_1$ = 201.9 MHz and $f_2$ = 214.1 MHz ($\Delta f$ = 12.2 MHz) under $\varphi_n = n\pi/3$ ($n \in [1, 6]$) phase configuration. Nondegenerate peaks correspond to clockwise and counterclockwise $k$-gap modes. **b**, Far-field validation of spectral splitting. Identical frequency separation at 1 m distance, confirming directional mode decoupling in radiation. **c**, Frequency-split intensity evolution. Measured $\Delta f$ versus $V_{mod}$. Splitting magnitude jumps >76 dBm above threshold ($V_{mod}$ > 0.44 V). **d**, Threshold-triggered amplification. Steady-state near-field (blue) and far-field (red) intensities versus $V_{mod}$. Step-like jump >43 dBm at 0.43 V confirms spatiotemporal PTC masing. **e-h**, Measured far-field intensity (e, g) and phase (g, h) distributions at 617.9 MHz (e-f) and 630.1 MHz (g-h), At 617.9 MHz, the unwrapped spiral phase distribution exhibits a $2\pi$ azimuthal progression, confirming a -1 topological charge. No helical phase characteristics are observed in the far field at 630.1 MHz.

To further verify far-field masing carrying OAMs in space-time modulated PTC metamaterials, we measure the far-field intensity and phase distributions (at the $z$=1 m plane) of the $E_\rho$ component at two split frequencies within the second Floquet sector, as shown in Figs. 4e-f for 617.9 MHz and Figs. 4g-h for 630.1 MHz. Consistent with theoretical predictions for discretized spatiotemporal modulation, mode coupling between CW and CCW components occurs in our system. This results in hybridization between CW and CCW modes at the two spectral peaks within each Floquet sector. At 617.9 MHz, the coupling between CW and CCW modes is relatively weak. Consequently, the dominant characteristics of the CW mode are preserved, evident in the toroidal intensity null at the beam center and the spiral phase distribution carrying a topological charge of approximately -1. The OAM beam exhibits a purity of 77.29%, which is comparable to that with other OAM generation methods (see S10 of the Supplementary Materials for details). However, at 630.1 MHz, the strong CW-CCW coupling suppresses the spiral phase distribution. These experimental results agree with the numerical simulations presented in Figs. 2l-m, confirming the experimental realization of coherent microwave emission carrying OAMs in our spatiotemporal PTCs. In S11 of the Supplementary Materials, we further compare the masing behavior between our model and conventional masers, showing unique features in terms of masing patterns, spectral characteristics and gain mechanism in our PTC masing.

**Discussion and Conclusion.** In conclusion, we have experimentally realized the first surface-emitted microwave lasing with OAMs through spatiotemporally engineered PTCs. In terms of theoretical innovation, we extend the PTC-lasing model from temporal modulation of three-dimensional bulk materials[15] to surface spatiotemporal modulation. This extension enables out-of-plane PTC lasing with OAMs. From the experimental perspective, we propose a time-varying capacitance design that differs from existing approaches. Unlike previous studies on PTCs that employed varactor diodes for temporal modulation[28–30], our design utilizes a multiplier-based time-varying capacitor, achieving nearly an order of magnitude enhancement in modulation strength. Thus, the generated $k$-gap can overcome both transmission and out-of-plane radiation losses, enabling the surface-emitted OAM masing.

    Moreover, PTC-enabled masing can efficiently generate high-frequency microwave emission using a low-frequency driving signal. This mechanism is significantly more efficient than traditional approaches relying on material nonlinearities[50] for high-frequency masing generation. Looking forward, our design strategy can be extended to higher frequencies, opening a viable pathway towards highly efficient terahertz masing sources. Equally significant is the spontaneous emergence of vortex-beam masing carrying OAMs, a direct manifestation of the symmetry breaking induced by the circularly configured space-time modulation. Unlike conventional masers, this platform encodes topological information directly into the masing mode—a capability with far-

reaching implications for structured-wavefront engineering. The demonstrated OAM generation mechanism aligns with the growing demand for compact, reconfigurable solutions in high-capacity communications and wave manipulation. These results establish a foundational framework for merging PTC physics with microwave lasing and active metamaterial engineering. More broadly, our work suggests that space-time engineered metamaterials could overcome fundamental limitations in photonic systems—from overcoming the laser linewidth limit via momentum-selective gain to enabling tunable masing/lasing frequencies. Future efforts to miniaturize the modulation architecture and explore higher-order OAM states may further unleash the potential of time-varying metamaterials.

**Note:** Upon completion of our work, we noted a recent study also reporting PTC-based microwave lasing[51]. This work utilized a varactor diode-modulated PTC structure confined within a closed waveguide. Unlike our PTC-based vertically emitting vortex maser, this configuration inherently precludes out-of-plane emission and the generation of structured maser fields.

**Reference**

1. Zhang, J. et al. Observation of a discrete time crystal. *Nature* **543**, 217 (2017).
2. Galiffi, E. et al. Photonics of time-varying media. *Adv. Photonics* **4**, 014002 (2022).
3. Engheta, N. Four-dimensional optics using time-varying metamaterials. *Science* **379**, 1190–1191 (2023).
4. Rudner, M. S., & Lindner, N. H. Band structure engineering and nonequilibrium dynamics in Floquet topological insulators. *Nat. Rev. Phys.* **2**, 229–244 (2020).
5. Yin, S., Galiffi, E. & Alù, A. Floquet metamaterials. *eLight* **2**, 8 (2022).
6. Yuan, L., Lin, Q., Xiao, M. & Fan, S. Synthetic dimension in photonics. *Optica* **5**, 1396 (2018).
7. Wang, Y. et al. Observation of Nonreciprocal Wave Propagation in a Dynamic Phononic Lattice. *Phys. Rev. Lett.* **121**, 194301 (2018).
8. Li, J. et al. Reciprocity of thermal diffusion in time-modulated systems. *Nat. Commun.* **13**, 167 (2022).
9. Holberg, D., & Kunz, K. Parametric properties of fields in a slab of time-varying permittivity. *IEEE Trans. Antennas Propag.* **14**, 183 (1966).
10. Biancalana, F., Amann, A., Uskov, A. V., & O'Reilly, E. P. Phys. Rev. E Stat. Nonlin. *Soft Matter Phys.* **75**, 046607 (2007).
11. Zurita-Sánchez, J. R., Halevi, P., & Cervantes-Gonzalez, J. C. Reflection and transmission of a wave incident on a slab with a time-periodic dielectric function ϵ(t). Phys. Rev. A 79, 053821 (2009).
12. Reyes-Ayona, J. R., & Halevi, P. Observation of genuine wave vector (k or β) gap in a dynamic transmission line and temporal photonic crystals. *Appl. Phys. Lett.* **107**, 074101 (2015).
13. Shaltout, A. M., Fang, J., Kildishev, A. V., & Shalaev, V. M. Photonic time-crystals and momentum band-gaps. *CLEO: QELS_Fundamental Science* FM1D-4 (2016).
14. Asgari, M. M., Garg P., Wang X., Mirmoosa M. S., Rockstuhl C., & Asadchy V. Theory and applications of photonic time crystals: A tutorial. *Adv. Opt. Photonics* **16**, 958 (2024).

15. Lyubarov, M., Lumer Y., Dikopoltsev A., Lustig E., Sharabi Y., & Segev M. Amplified emission and lasing in photonic time crystals. *Science* **377**, 425 (2022).
16. Lustig, E., Sharabi, Y., & Segev, M. Topological aspects of photonic time crystals. *Optica* **5**, 1390–1395 (2018).
17. Ren, Y. et al. Observation of momentum-gap topology of light at temporal interfaces in a time-synthetic lattice. *Nat. Commun*. **16**, 707 (2025).
18. Feis, J., Weidemann, S., Sheppard, T. et al. Space-time-topological events in photonic quantum walks. *Nat. Photon.* **19**, 518–525 (2025).
19. Dikopoltsev, A. et al. Light emission by free electrons in photonic time-crystals. *Proc. Natl. Acad. Sci. U.S.A.* **119**, e2119705119 (2022).
20. Li, H., Yin, S., He, H., Xu, J., Alù, A. & Shapiro, B. Shapiro, Stationary charge radiation in anisotropic photonic time crystals. *Phys. Rev. Lett.* **130**, 093803 (2023)
21. Pan, Y., Cohen, M. I. & Segev, M. Superluminal k-gap solitons in nonlinear photonic time crystals, *Phys. Rev. Lett.* **130**, 233801 (2023).
22. Wang, X., Garg, P., Mirmoosa, M.S. et al. Expanding momentum bandgaps in photonic time crystals through resonances. *Nat. Photon.* **19**, 149–155 (2025).
23. Sharabi, Y., Lustig, E., & Segev, M. Disordered photonic time crystals. *Phys. Rev. Lett.* **126**, 163902 (2021)
24. Sharabi, Y., Dikopoltsev, A., Lustig, E., Lumer, Y., & Segev, M. Spatiotemporal photonic crystals. *Optica* **9**, 585-592 (2022).
25. Yousefjani, R., Sacha, K., & Bayat, A. Discrete time crystal phase as a resource for quantum-enhanced sensing. *Phys. Rev. B* **111**, 125159 (2025).
26. Huidobro, P. A., Galiffi, E., Guenneau, S., Craster, R. V., & Pendry, J. B. Fresnel drag in space–time-modulated metamaterials. *Proc. Nat. Acad. Sci.* **116**, 24943-24948 (2019).
27. Galiffi, E., Huidobro, P. A., & Pendry, J. B. Broadband nonreciprocal amplification in luminal metamaterials. *Phys. Rev. Lett.* **123**, 206101 (2019).
28. Park, J., Cho, H., Lee, S. et al. Revealing non-Hermitian band structure of photonic floquet media. *Sci. Adv.* **8**, eabo6220 (2022).
29. Ye, X., Wang, Y. G., Yao, J. F., Wang, Y., Yuan, C. X., & Zhou, Z. X. Realization of Spatiotemporal Photonic Crystals Based on Active Metasurface. *Laser & Photonics Rev.* **19**, 2401345 (2025).
30. Wang, X. et al. Metasurface-based realization of photonic time crystals. *Sci. Adv.* **9**, eadg7541 (2023).
31. Gordon, J. P., Zeiger, H. J., & Townes, C. H. Molecular microwave oscillator and new hyperfine structure in the microwave spectrum of NH3. *Phys. Rev.*, **95**, 282–284 (1954).
32. Maiman, T. H. Stimulated optical radiation in ruby. *Nature* **187**, 493–494 (1960).
33. Heard, H. G. Ultra-violet gas laser at room temperature, *Nature* **200**, 667–668 (1963).
34. Li, C. H., Benedick A. J., Fendel, P., Glenday, A. G., Kärtner, F. X., Phillips, D. F., Sasselov, D., Szentgyorgyi, A., & Walsworth, R. L. A laser frequency comb that enables radial velocity measurements with a precision of 1cm s$^{-1}$. *Nature* **452**, 610–612 (2008).
35. Hecht, J. Short history of laser development. *Appl. Opt.* **49**, F99 (2010).
36. Makhov, G., Kikuchi, C., Lambe, J., & Terhune, R. W. Maser action in ruby. *Phys. Rev.* **109**, 1399 (1958).
37. Siegman, A. E., & Hagger, H. J. Microwave Solid-state Masers (1964).
38. Joshin, K., Mimura T., Ninori, M.,Yamashita Y., Kosemura, K., & Saito, J. Noise performance of microwave HEMT. *IEEE MTT-S International Microwave Symposium Digest*, 563–565 (1983).
39. Oxborrow, M., Breeze J. D., & Alford N. M. Room temperature solid-state maser. *Nature* **488**, 353 (2012).
40. Breeze, J. D., Salvadori E., Sathian, J., Alford N. M., & Kay C. W. Continuous-wave room-temperature diamond maser. *Nature* **555**, 493 (2018).
41. Day, T., Isarov, M., Pappas, W. J., Johnson, et al. Room-temperature solid-state maser

amplifier. *Phys. Rev. X* **14**, 041066 (2024).
42. Jin, L., Pfender, M., Aslam, N., Neumann, P., Yang, S., Wrachtrup, J., & Liu, R. B. Proposal for a room-temperature diamond maser. *Nat. Commun.* **6**, 8251(2015).
43. Forbes, A., Mkhumbuza, L. & Feng, L. Orbital angular momentum lasers. *Nat. Rev. Phys.* **6**, 352–364 (2024).
44. Wright, L. G., Christodoulides, D. N., & Wise, F. W. Spatiotemporal mode-locking in multimode fiber lasers. *Science* **358**, 94 (2017).
45. Malomed, B. A., Mihalache, D., Wise, F., & Torner, L. Spatiotemporal optical solitons. *J. Opt. B: Quantum Semiclass. Opt.* **7**, R53 (2005).
46. Wright, L. G., Renninger, W. H., Christodoulides, D. N., & Wise, F. W. Spatiotemporal dynamics of multimode optical solitons. *Opt. Express* **23**, 3492 (2015).
47. Haus, H. A. Mode-locking of lasers. *IEEE J. Sel. Top. Quantum Electron.* **6**, 1173 (2002).
48. Cao, B., Gao, C., Liu, K., Xiao, X., Yang, C., & Bao, C. Spatiotemporal mode-locking and dissipative solitons in multimode fiber lasers. *Light Sci. Appl.* **12**, 260 (2023).
49. Cai, X. et al., Integrated compact optical vortex beam emitters. *Science* **338**, 363–366 (2012)
50. Lu, Y. et al. Nonlinear optical physics at terahertz frequency. *Nanophotonics* **13**, 3279-3298 (2024).
51. Lee, K. et al., Spontaneous emission and lasing in photonic time crystals. *arXiv preprint arXiv*: 2507.19916 (2025).

**Acknowledgements.** This work is supported by the National Key R & D Program of China under Grant No. 2022YFA1404900, National Natural Science Foundation of China No. 12422411, Beijing Natural Science Foundation No. 1242027 and Young Elite Scientists Sponsorship Program by CAST No. 2023QNRC001.

**Methods**

**I. Full-wave simulations of PTCs based on finite element methods.**

All numerical simulations are performed using the finite element software COMSOL Multiphysics. The transient (temw) and frequency domain (emw) modules are employed to analyze wave evolutions of PTCs and S-parameters of the model, respectively.

For the idealized PTC model, a time-varying surface capacitance is implemented using the surface current density boundary condition, defined as $\bm{J} = d(C\bm{E})/dt$, where the capacitance $C = C_0(1 + \delta sin(2\omega t))$ with parameters $C_0$=0.5 pF, $\delta$=0.1 and $\omega/2\pi$=317 MHz. This boundary condition is applied to a ring-shaped region with a width of $g$=1 mm and radius $r$=0.1m, while the remaining areas of the ring plane are modeled as perfect electric conductor boundaries. To investigate the masing behavior induced by $k$-gap, the time-domain simulations are conducted in two stages. The **initial excitation phase ($t$ = 0 to 10T):** A line current source with the form $I(t) = \sin(\omega t)$ is applied along a radial section of the ring to excite the static model with $\delta$=0. Then, the **modulation phase ($t$ > 10T):** After removing the excitation source at $t$=10T, the time-varying capacitance modulation is activated to observe the temporal evolution of electric fields in PTC. During this phase, the electromagnetic signal exhibits exponential amplification due to the $k$-gap.

To realistically model saturation effects, the modulation depth $\delta$ was dynamically adjusted as a function of the field intensity, decreasing as the field strength increased. For the spatiotemporally modulated surface capacitance model, the same procedure is followed, with the capacitance expression modified to $C = C_0(1 + \delta sin(2\omega t + q\varphi))$.

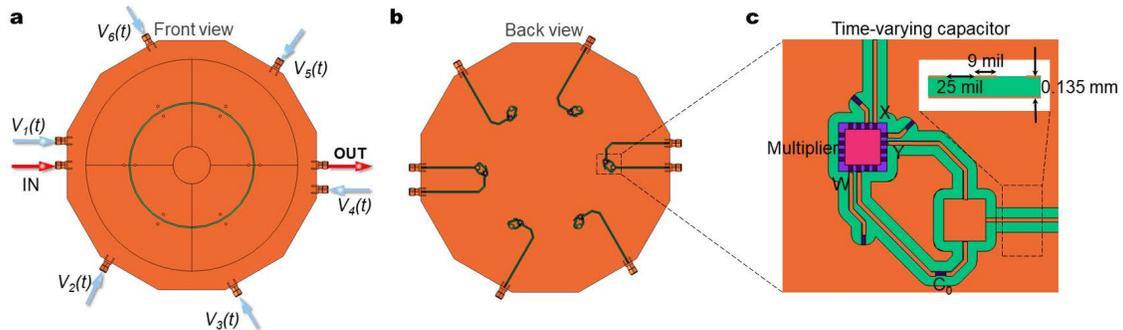

**Figure 5 | The schematic diagram of time-varying metamaterials for implementing PTCs.** Schematic illustrations of the simulated PTC structure: **a,** front view, and **b,** rear view of the ring-shaped microstrip with integrated multiplier-driven time-varying capacitance units. **c,** enlarged view of a single unit (dashed box in **c**), highlighting the feedline design (width:9 mil height above ground: 0.135 mm) for 50 Ω impedance matching. The multiplier (ADL5391) is modeled using voltage-controlled sources with input impedance 200 Ω and output clamped to ±1 V, connected via a 20 nH inductor.

Next, we detail the method for the practical structure simulations. To simulate the multiplier-driven time-varying capacitance and its coupling with the electromagnetic planar waveguide, we combine the electromagnetic transient (*temw*) and circuit (*cir*) modules. The coupling between the electromagnetic field and the circuit is achieved using the "External I vs. U" interface in the lumped port boundary condition. The practical structure is illustrated in Figs. 5a-c, with Figs. 5a-b showing the front and rear views, respectively. Fig. 5c provides a magnified view of the multiplier-based time-varying capacitance region, including the feedline design (width: 9 mil; height above the ground substrate: 0.135 mm), which ensured a characteristic impedance of 50 Ω for impedance matching with the SMA ports. The substrate material was S1000HB (relative permittivity $\varepsilon_r$=2.4). In addition, since the ADL5391 multiplier lacked a COMSOL-compatible SPICE netlist, we built a multiplier with saturation effect using voltage-controlled voltage sources, and its input port(X,Y) impedance is set to 200 Ω based on the datasheet and experimental measurements, with a 20 nH inductor connected in series at the output port(W), as shown in Fig. 5c. The input and output voltages of multipliers are clamped to ±1 V to reflect the device's operational limits. The input (red arrow in Fig. 5b) and modulation (blue arrows) signals are fed into the structure via 50 Ω SMA ports.

In addition, the frequency-domain (*emw*) module is used to calculate the S-parameters of the unmodulated structure. The same geometry is employed, with circuit elements replaced by lumped element boundary conditions. The transmission spectrum is obtained by exciting the input port and measuring the response at the output port (see Fig. S4).

**II. Far-field calculations using angular spectrum theory.**

The angular spectrum theory is employed to compute far-field radiation patterns from the numerically calculated near-field distributions. The workflow is as follows. First, the near-field electromagnetic field distribution $U(x, y, 0, t)$ is obtained through simulation via COMSOL. Performing a Fourier transform along the time dimension yields the frequency-domain field distribution $U(x, y, 0, \omega)$ at each spatial point. Then, the angular spectrum method is employed to model the free-space propagation of scalar electromagnetic fields. Under the paraxial or non-paraxial approximation, an arbitrary complex optical field $U(x, y, 0, \omega)$ defined at the source plane can be decomposed into a superposition of plane waves by computing its spatial Fourier transform $\tilde{U}(k_x, k_y, 0, \omega) = \mathcal{F}\{U(x, y, 0, \omega)\}$, where $\tilde{U}(k_x, k_y, 0, \omega)$ is the angular spectrum of the field, and $(k_x, k_y)$ are the transverse components of the wavevector. To achieve higher accuracy in far-field distribution, zero-padding is applied to the periphery of the source-plane field distribution $U(x, y, 0, \omega)$ obtained from simulations. The propagation of each spectral component over a distance $z$ in free space introduces a phase delay given by $\tilde{U}(k_x, k_y, z, \omega) = \tilde{U}(k_x, k_y, 0, \omega) \cdot exp(ik_z z)$, where $k_z = \sqrt{k^2 - k_x^2 - k_y^2}$, and $k = \omega/c$ is the wavenumber in vacuum. For evanescent components $(k_x^2 + k_y^2 > k^2)$, $k_z$ becomes imaginary, representing exponential decay. The propagated field $U(x, y, z, \omega)$ at distance $z$ is then obtained by the inverse Fourier transform $U(k_x, k_y, z, \omega) = \mathcal{F}^{-1}\{\tilde{U}(x, y, z, \omega)\}$. This method ensures accurate modeling of near- and far-field diffraction without resorting to approximations such as the Fresnel or Fraunhofer limits. It is particularly suitable for numerical implementations using the fast Fourier transform, enabling efficient simulation of wavefront propagation in free space or through homogeneous media.

**III. Detailed derivation of multiplier-driven time-varying capacitance.**

In this part, we provide the detailed derivation on the time-varying capacitance circuit governed by multiplier, whose input-output relations is expressed as $W = \alpha XY$, with *X*, *Y* being input voltages, *W* being the output voltage, and $\alpha$ being a scaling factor adjustable between 0.5 and 2.5 via external DC voltage. One input port connects directly to the top planar waveguide signal $U(t)$, while the other receives the external modulation signal $V(t)$. A fixed capacitor $C_0$ links the multiplier output to the planar waveguide. Assuming the feedline length is negligible compared to the wavelength, voltage fluctuations can be ignored. Thus, the current through $C_0$ is $i_1 = C_0 \left[ (1 - \alpha V(t)) \frac{dU(t)}{dt} - \alpha U \frac{dV(t)}{dt} \right]$. The total current $i$ flowing into time-varying capacitor from the transmission line splits into $i_1$ (flowing through $C_0$) and $i_2$ (entering the multiplier input). Due to the high input impedance, $i_1 \gg i_2$, so $i \approx i_1$. The equivalent time-varying capacitance $C(t)$ satisfies $i = \frac{d(C(t)U(t))}{dt} = C(t)\frac{d(U(t))}{dt} + U(t)\frac{d(C(t))}{dt}$. Substituting $i_1$ yields: $C_0 \left[ (1 - \alpha V(t)) \frac{dU(t)}{dt} - \alpha U(t) \frac{dV(t)}{dt} \right] = C(t)\frac{d(U(t))}{dt} + U(t)\frac{d(C(t))}{dt}$. Solving this gives $C(t) = C_0(1 - \alpha V(t))$. Thus, the equivalent capacitance is directly controlled by $V(t)$.

**IV. Sample Fabrication and Measurement Methods.**

The time-varying capacitance and transmission line modules are designed using circuit design software and fabricated separately. Post-fabrication, interconnects were established using metal conduits. Time-Varying Capacitance Board: A four-layer design with copper foils (L1–L4) separated by three dielectric layers. Outer layers (L1, L4) housed circuit traces and components, while inner layers (L2, L3) served as ground planes. To minimize crosstalk, feedlines and components were shielded by 25 mil (0.635 mm) copper barriers connected to the ground via arrays. The outer dielectric (thickness: 0.1355 mm) was optimized for microwave transmission, while a thicker middle layer provided mechanical support. Transmission Line: A two-layer structure with a single 2.4 mm dielectric layer, featuring copper-clad edges for interlayer connectivity.

Instrumentation included a four-channel signal generator, spectrum analyzer, oscilloscope, network analyzer, loop antenna, coaxial cables, and multi-output DC power supplies. Temporal Modulation: Three synchronized 416 MHz signals (0.5 V amplitude, differential ± outputs) were routed to three multipliers. A fourth signal (208 MHz, 0.025 V) excited the input port, while the output was monitored via a spectrum analyzer. The multipliers were powered by +5 V DC, with bias voltages (2.0–2.8 V) tuning $\alpha$ for optimal performance. Removing the 208 MHz input confirmed microwave lasing, as the output signal persisted due to gain exceeding losses. Near-/Far-Field Characterization: Near-field intensity versus modulation voltage (0.05–0.5 V) was measured directly. Far-field measurements used a loop antenna at 1 m distance. Spatiotemporal Modulation: The negative outputs of the three 416 MHz signals were connected to the remaining multipliers. Far-field amplitude and phase profiles were measured using an oscilloscope, with phase resolved via Fourier analysis of reference and scanned signals.